\documentclass[twocolumn,aps,prl,citeautoscript,showpacs]{revtex4-1}
\pdfoutput=1

\usepackage{amsmath}
\usepackage{amssymb}
\usepackage{graphicx}
\usepackage{hyperref}

\newcommand{\abs}[1]{\left| #1 \right|}
\renewcommand{\vec}[1]{\mathbf{#1}}
\newcommand{\eps}{\varepsilon}
\newcommand{\vE}{\vec{E}}
\newcommand{\vG}{\vec{G}}

\newcommand{\vk}{\vec{k}}

\newcommand{\vq}{\vec{q}}
\newcommand{\vu}{\vec{u}}
\newcommand{\vv}{\vec{v}}
\newcommand{\pd}[2]{\ensuremath{\frac{\partial #1}{\partial #2}}}
\begin{document}

\title{Conditions for $T^2$ resistivity from electron-electron scattering}

\author{Michael Swift}
\affiliation{Department of Physics, University of California, Santa Barbara, California 93106-9530, USA}

\author{Chris G. Van de Walle}
\affiliation{Materials Department, University of California, Santa Barbara, California 93106-5050, USA}

\begin{abstract}
Many complex oxides (including titanates, nickelates and cuprates) show a regime in which resistivity follows a power law in temperature ($\rho\propto T^2$).  By analogy to a similar phenomenon observed in some metals at low temperature, this has often been attributed to electron-electron (Baber) scattering.  We show that Baber scattering results in a $T^2$ power law only under several crucial assumptions which may not hold for complex oxides.  We illustrate this with sodium metal ($\rho_\text{el-el}\propto T^2$) and strontium titanate ($\rho_\text{el-el}\not\propto T^2$).  We conclude that an observation of $\rho\propto T^2$ is not sufficient evidence for electron-electron scattering.
\end{abstract}

\maketitle


A resistivity component of the form $\rho = A T^2$ has been observed in a variety of materials.  The most well known are metals at low temperature, including transition metals~\cite{White59,White67,Anderson68} (with $A\sim 10^{-4} \text{ to } 10^{-5} \,\mu \Omega\, \text{cm} / \text{K}^2$) and alkali metals~\cite{Bass90,vanKempen76,Levy79} (with $A\sim10^{-6}\text{ to }10^{-7}\,\mu \Omega\, \text{cm} / \text{K}^2$).  The mechanism behind this contribution to resistivity has been identified as electron-electron scattering (or Baber scattering~\cite{Baber37}).  This scattering mechanism is well described by Fermi liquid theory, which predicts $\rho_\text{el-el}=A T^2$ of a similar magnitude to that seen in experiments~\cite{Rice68,Lawrence73,MacDonald81}.  Unifying features of electron-electron scattering in these materials include a relatively small scattering rate and a low temperature threshold ($\sim 20$ K for transition metals, a few K for alkali metals) above which other scattering mechanisms (such as electron-phonon) dominate.

More recently, investigations into transport properties of complex oxides have also found a resistivity component $\rho = A T^2$, or a component of electron mobility $\mu = \alpha T^{-2}$.  Examples include SrTiO$_3$ (STO),~\cite{Marel11,Mikheev15APL,Lin15,Mikheev16} rare-earth nickelates~\cite{Zhou05,Mikheev15sciadv}, and cuprates~\cite{Ando04,Barisic13}.  Discussions of the $T^2$ behavior (and deviations from it) are often based on the assumption that the $T^2$ comes from Fermi-liquid electron-electron scattering, much as it does in metals.  Though the power law is the same, this mechanism is several orders of magnitude stronger ($A\sim 10^{-1} \text{ to } 10^{-2} \,\mu \Omega\, \text{cm} / \text{K}^2$) and has been observed in some cases up to room temperature.  Measurements of other transport signatures have also clashed with predictions from Fermi liquid theory~\cite{Mikheev16}.

In this work, we show that the $T^2$ exponent of Baber scattering arises only under a certain set of assumptions.  These assumptions are fulfilled in metals at low temperature (as we show explicitly for the case of sodium metal), but are not necessarily fulfilled in semiconductors at higher temperature.  In the specific case of bulk STO, we find that many of the assumptions necessary to observe $T^2$ via Baber scattering do not hold, and explicit calculations of this scattering mechanism result in a resistivity that significantly deviates from the $T^2$ behavior.  More generally, our findings imply that observation of $\rho\propto T^2$ should not be treated as a ``smoking gun'' for electron-electron scattering, and more careful analysis is needed to establish specific mechanisms for a given system.


We approach the study of this electron-electron scattering using Boltzmann transport theory, following methods derived in Refs.~\onlinecite{Ziman60,Klimin12}.
The equilibrium occupation of a state in band $n$ with crystal momentum $\vk$ and energy $\eps_{n,\vk}$ is given by the Fermi-Dirac distribution
\begin{equation}\label{f-definition}
f(\eps_{n,\vk}) =  \left(\exp\left(\frac{\eps_{n,\vk}-\mu}{k_B T}\right)+1\right)^{-1} \, ,
\end{equation}
where $\mu$ is the chemical potential, $T$ is the temperature, and $k_B$ is the Boltzmann constant.  $\eps_{n,\vk} $ and $\mu$ are referenced to the conduction-band minimum.  The Boltzmann transport equation describes the effects of external forces, diffusion, and internal collisions on the time evolution of the distribution function:
\begin{equation}\label{Bolzmann-eq}
\vv\cdot\nabla f + \frac{e}{\hbar}\vE\cdot\pd{f}{\vk} = \left(\pd{f}{t}\right)_\text{scatt} \, .
\end{equation}

We introduce $\Phi_{n,\vk}$, the deviation of the distribution function from equilibrium:
\begin{equation}
f_{n,\vk} = f(\eps_{n,\vk}) + \Phi_{n,\vk}\pd{f(\eps)}{\eps} \, .
\end{equation}
Letting $X$ be the left-hand side of Eq.~(\ref{Bolzmann-eq}) and $P$ be an operator representing the effects of scattering, it can be shown \cite{Ziman60} that Eq.~(\ref{Bolzmann-eq}) may be written as $X =  P \Phi$.  With the inner product $\langle A, B\rangle = \sum_n \int d\vk\, AB $, this implies
\begin{equation}
\langle \Phi, X\rangle  =  \langle \Phi, P \Phi\rangle \, .
\end{equation}
The variational principle established by Ziman in Ref.~\onlinecite{Ziman60} shows that the solution $\Phi$ minimizes $\langle \Phi, P \Phi\rangle$.  Using the trial function $\Phi=\vv\cdot\hat\vu$, where $\hat\vu$ is the vector direction of the electric field, this gives a collision integral for electron-electron scattering~\cite{Klimin12},
\begin{align}\label{coll-int}
\langle \Phi, & P\Phi\rangle = \ \frac{1}{2 k_B T}\frac{1}{(2\pi)^9}\frac{2\pi}{\hbar}\sum_{n,n'}\int d\vk_1 d\vk_2 d\vk_3 d\vk_4 \\
\notag & \times \left[\left(\vv_{n,\vk_1}+\vv_{n',\vk_2}-\vv_{n,\vk_3}-\vv_{n',\vk_4}\right)\cdot \hat\vu\right]^2\\
\notag & \times f\left(\eps_{n,\vk_1}\right)f\left(\eps_{n',\vk_2}\right)\left[1-f\left(\eps_{n,\vk_3}\right)\right]\left[1-f\left(\eps_{n',\vk_4}\right)\right]\\
\notag & \times \left(U_{\vk_1,\vk_3}^\text{(eff)}\right)^2 \delta\left(\eps_{n,\vk_1}+\eps_{n',\vk_2}-\eps_{n,\vk_3}-\eps_{n',\vk_4}\right)\\
\notag & \times \delta\left(\vk_1+\vk_2-\vk_3-\vk_4\right) \, .
\end{align}
The integral takes into account scattering between two electrons in bands $n$ and $n'$ with initial momenta $\vk_1$ and $\vk_2$ and final momenta $\vk_3$ and $\vk_4$.
$U_{\vk_1,\vk_3}^\text{(eff)}$ is the effective interaction for the momentum transfer $\vk_3-\vk_1$, and ${\vv_{n,\vk} = 1/\hbar\ \partial\eps/\partial\vk}$ is the band velocity.  Note that momentum conservation sends the velocity term (and thus the entire expression) to zero in the absence of a mechanism for ``momentum relaxation''.  This can be provided by Umklapp processes ($\vk_1+\vk_2-\vk_3-\vk_4 = \vG$, a lattice vector) or by scattering between states with different masses~\cite{Pal12}.

Equation~\ref{coll-int} can be normalized to give the resistivity:
\begin{align}
\label{rho-def}
\rho_\text{el-el} & = \mathcal{N}\langle \Phi, P\Phi\rangle \, , \, \text{with} \ \mathcal{N} = \left[\langle\Phi,X(E=1)\rangle\right]^{-2}\, ,
\end{align}
where $X(E=1)$ indicates the left-hand side of the Boltzmann transport equation with a unit electric field.  The normalization factor $\mathcal{N}$ is given by
\begin{equation}
\mathcal{N} = \frac{(2\pi)^6}{4e^2} \left[\sum_n\int d\vk \vv_{n,\vk}\Phi_{n,\vk}\pd{f(\eps_{n,vk})}{\eps_{n,\vk}}\right]^{-2} \, .
\end{equation}


We now lay out the standard derivation of a $T^2$ power law from Eq.~(\ref{coll-int}), in order to understand the assumptions involved.  We begin by separating the $\vk$-space integral into an integral over the Fermi surface and an integral in the perpendicular direction.  Assuming the chemical potential is constant with temperature (Assumption 1), integration over the Fermi surface will give a result that is independent of temperature.  In the direction perpendicular to the Fermi surface, the Fermi function terms and energy and momentum conservation restrict the scattering states to a narrow thermal envelope around the Fermi surface.  Assuming the non-Fermi-function terms vary slowly enough over the width of this envelope (Assumption 2), they may be approximated as constants given by their value at the Fermi surface.  Ignoring the temperature-independent terms and changing integration variables to energy, we can define the integral $I$, which contains the temperature dependence of the resistivity:
\begin{align}\label{rho_fermi}
\rho_\text{el-el} \propto I =\ & \frac{3}{2\pi^2 k_B^3 T} \int_0^\infty d\eps_1\,d\eps_2\,d\eps_3\, f\left(\eps_1\right)f\left(\eps_2\right)\\
& \notag \times \left[1-f\left(\eps_3\right)\right]\left[1-f\left(\eps_1+\eps_2-\eps_3\right)\right] \,\end{align}
We extract the dimensionful quantities by changing integration variables again to $x_i=(\eps_i-\mu)/k_BT$ and assuming the lower bound on $x_i$ integration can be taken to $-\infty$ (Assumption 3):
\begin{align}\label{I}
I=\ & \frac{3T^2 }{2\pi^2} \int_{-\infty}^\infty dx_1\,dx_2\,dx_3\, (e^{x_1}+1)^{-1}(e^{x_2}+1)^{-1}\\
\notag & \times \left[1-(e^{x_3}+1)^{-1}\right]\left[1-(e^{x_1+x_2-x_3}+1)^{-1}\right]\,
\end{align}
The integral is a dimensionless constant, so $\rho_\text{el-el}\propto T^2$.

As highlighted during the course of the derivation, this result depends on three key assumptions.
We will explore each of the assumptions and determine whether they are satisfied in our case studies, Na metal and STO.  We will investigate the impact of each assumption on the final result by calculating $\rho$ as a function of $T$ employing the assumption and comparing to a numerical calculation with the assumption removed.  Numerical integration is carried out with Divonne, a Monte Carlo integration algorithm which uses stratified sampling for variance reduction, as implemented in the \textsc{Cuba} package~\cite{Hahn05}.

\begin{figure}
\includegraphics[width=8.6cm]{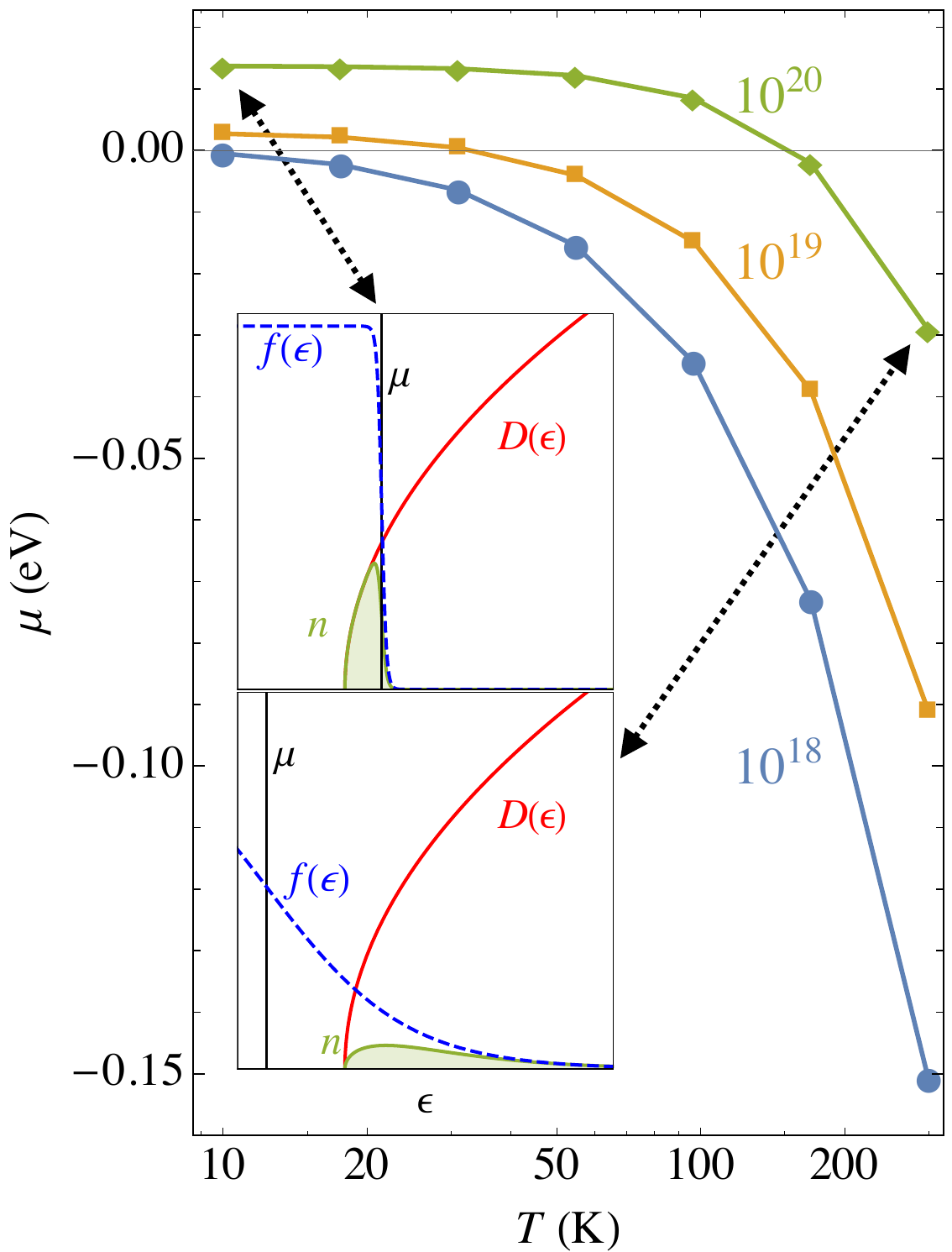}
\caption{Chemical potential as a function of temperature in strontium titanate for three different fixed electron concentrations, expressed in cm$^{-3}$.  Inset: illustration of Eq.~(\ref{find-mu}) applied to find $\mu$ for $n=10^{20}$ cm$^{-3}$ at $T=10$ K and $T=300$ K.  These results were calculated using our anisotropic parabolic fit to the conduction band of SrTiO$_3$ [Eq.~(\ref{STO_dispersion})].}
\label{muvsT}
\end{figure}

\noindent \underline{Assumption 1}: As temperature changes, the chemical potential $\mu$ of the electrons is constant and equal to its zero-temperature value.  While this is a very good approximation in metals at low temperature, it may not hold in semiconductors at intermediate temperature.  In many cases (including degenerately doped semiconductors), the quantity which actually remains constant with temperature is the electron density $n$.  The chemical potential $\mu$ is determined by the equation
\begin{equation}\label{find-mu}
n = \sum_{i}\int d\eps\ D_i(\eps)f(\eps) \,,
\end{equation}
where $D_i$ is the density of states of band $i$, and $\mu$ and $T$ are implicit in $f$ [Eq.~(\ref{f-definition})].  With Assumption 1 in place, $\mu$ is set to its zero-temperature value, which we calculate analytically.  When this assumption is relaxed, the integral in Eq.~(\ref{find-mu}) is calculated numerically, and $\mu$ is recalculated at any given temperature to keep $n$ fixed.  When $\mu \gg k_B T$,
smearing of $f$ due to increased $T$ does not have a strong effect on Eq.~(\ref{find-mu}),
so $\mu$ has negligible temperature dependence.
However, when $\mu\sim k_B T$, the chemical potential does have a significant temperature dependence, as shown in Fig.~\ref{muvsT}.

\noindent \underline{Assumption 2}: The integrand in Eq.~(\ref{coll-int}) is slowly varying compared to the Fermi functions over the width of the thermal envelope.  When Assumption 2 is in place, the non-Fermi-function terms in the integrand are taken to be constant for a given direction in $k$-space, equal to their value at the Fermi surface.  This reduces the radial part of the integral to Eq.~(\ref{rho_fermi}).  When Assumption 2 is relaxed, the radial dependence of the integrand is included explicitly.  This assumption is valid at temperatures that are small compared to the scale over which the non-Fermi-function terms vary.  This scale is difficult to predict \emph{a priori}, so we will assess the validity of this assumption on a case-by-case basis.

\begin{figure}
\includegraphics[width=8.6cm]{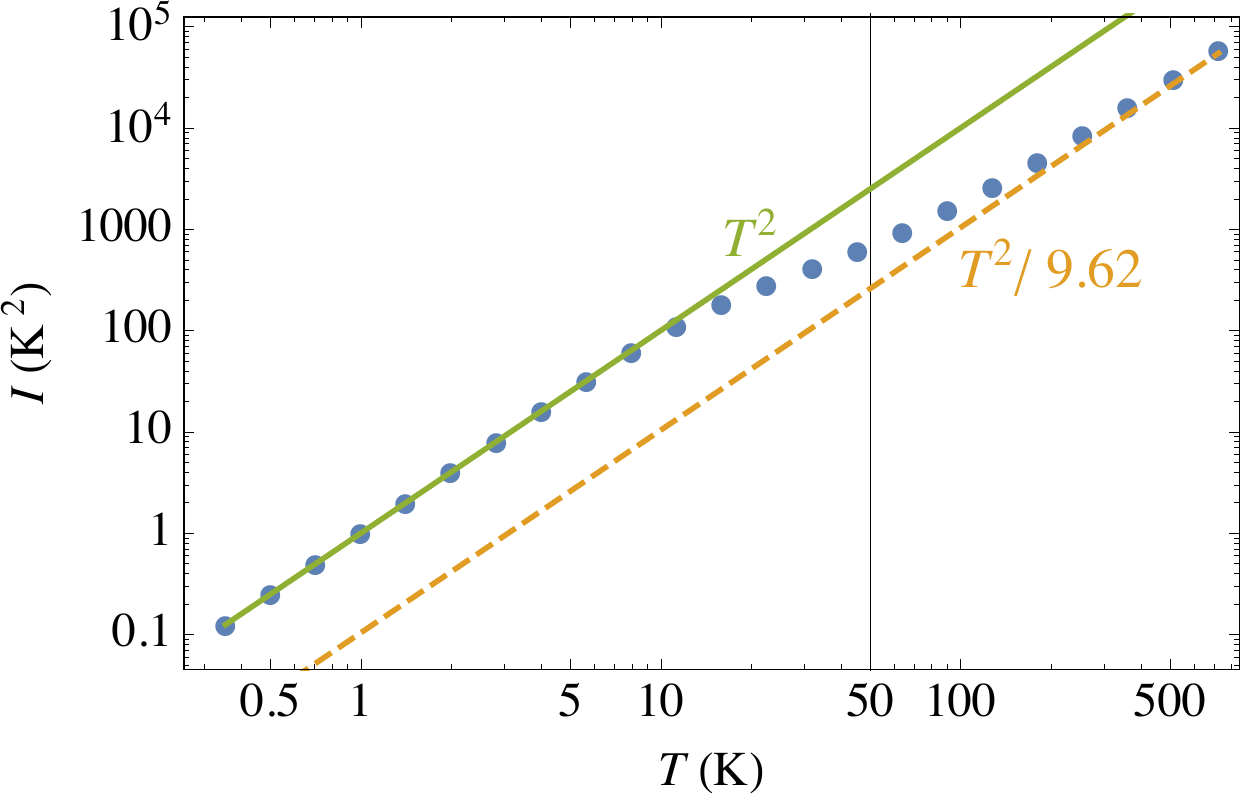}
\caption{Log-log plot of $I$ [Eq.~(\ref{I})] as a function of temperature for fixed chemical potential $\mu/k_B=50$ K (indicated by the solid vertical line).  $I$ shows a clear change from $T^2$ to $T^2/9.62$.}
\label{IvsTfig}
\end{figure}

\noindent \underline{Assumption 3}: The lower limit of integration in Eq.~(\ref{I}) can be taken to $-\infty$.  In fact, since the lower limit in Eq.~(\ref{rho-def}) is set by the conduction-band minimum ($\eps=0$), the lower limit of integral $I$  should be $-\mu/k_B T$.  If $\mu \gg k_BT$, Assumption 3 is valid and $I=T^2$.  However, if $\mu \ll k_BT $, Eq.~(\ref{I}) should instead run from 0 to $\infty$, and ${I = T^2(1/4 - 3(\ln 2)^2/\pi^2) \approx T^2/9.62}$.
These different prefactors imply that an intermediate regime ($\mu \sim k_B T$) must exist, in which $I\not\propto T^2$.  This becomes obvious when evaluating $I$ versus $T$ numerically, as shown in Fig.~\ref{IvsTfig}.

\begin{figure}
\includegraphics[width=8.6cm]{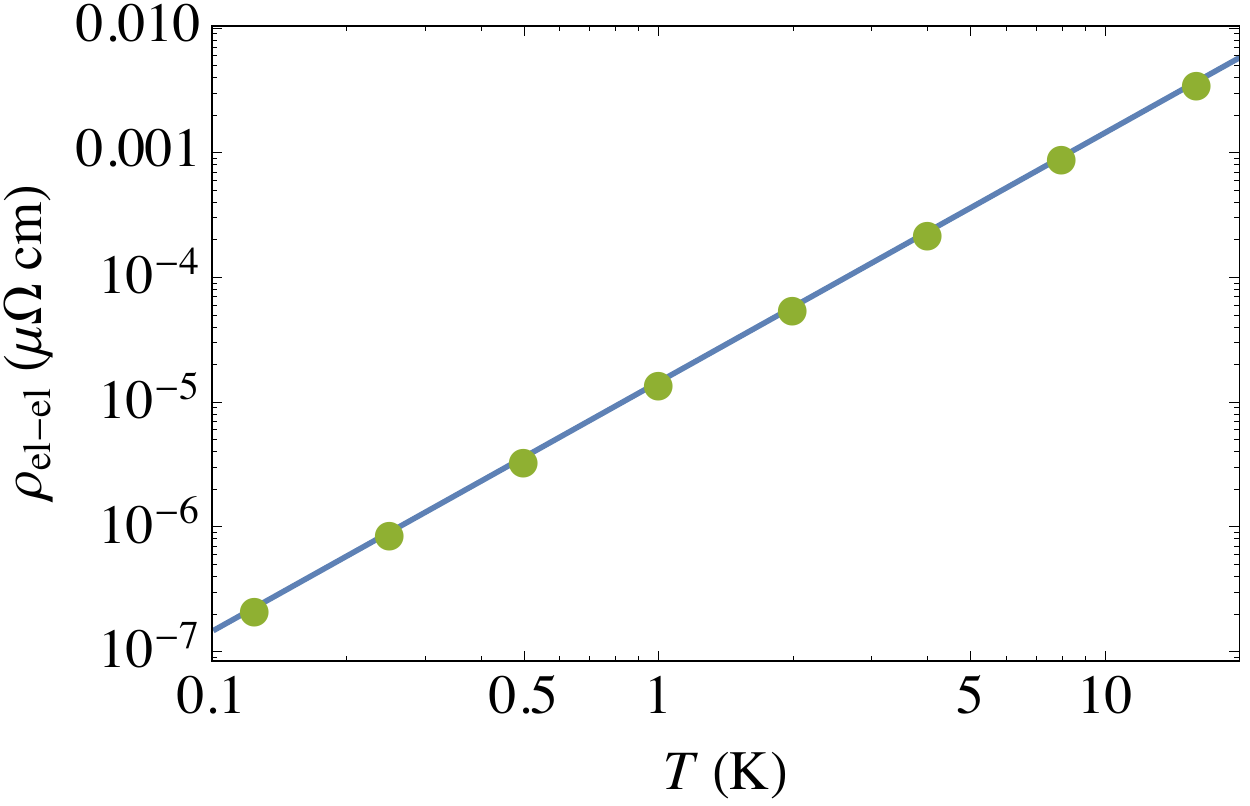}
\caption{Resistivity due to electron-electron scattering versus temperature for metallic sodium.  The solid blue line was calculated using all the assumptions while the green data points were calculated without any assumptions.  The assumptions hold in this case, so the two computations agree.}
\label{Na-fig}
\end{figure}

We now proceed to apply this methodology, starting with the test case of sodium metal.  We model the band structure by first performing a density functional theory calculation (as implemented in the Vienna Ab initio Simulation Package ({\sc VASP})~\cite{VASP}, employing the Perdew, Burke and Ernzerhof~\cite{PBE} functional), then fitting the resulting conduction band to a parabolic dispersion relation.  We find a parabolic effective mass of $m=1.06\, m_e$ produces a good fit with an accuracy better than 0.05 eV compared to the first-principles result.  Since sodium has a single parabolic band, momentum relaxation comes from Umklapp scattering.  The effective interaction between electrons is the screened Coulomb interaction
\begin{equation}
U_{\vq}^\text{(eff)} = \frac{4\pi e^2}{q^2 + \kappa^2} \, ,
\end{equation}
where $\vq=\vk-\vk'$ and $\kappa^2=4\pi e^2 \pd{n}{\mu}$ is the Lindhard screening length.
The zero-temperature chemical potential is $\mu=1.96$ eV.  This guarantees $\mu \gg k_B T$, so we expect the chemical potential to stay fairly constant with temperature (Assumption 1), and we expect to be in the regime where $I=T^2$ (Assumption 3).  Additionally, since the scattering mechanism is usually observed at very low $T$, the thermal envelope is quite narrow, so we would expect Assumption 2 to hold as well.  Our numerical results confirm these expectations (Fig. ~\ref{Na-fig}).  All the assumptions hold well in sodium, so $\rho_\text{el-el} \propto T^2$.

It is worth noting that the prefactor $A=1.4\times 10^{-5}\,\mu \Omega\, \text{cm} / \text{K}^2$ is larger than that observed experimentally ($A=1.7-2.2\times10^{-6}\,\mu \Omega\, \text{cm} / \text{K}^2$~\cite{Bass90}).  A careful inclusion of phonon-mediated electron-electron scattering could bring the calculated value closer to experiment~\cite{MacDonald80}.  This mechanism will not invalidate the assumptions or change the $T^2$ power law, so we do not discuss it further here.


We now turn to electron-electron scattering in doped STO.  We consider doping levels and temperatures that correspond to experimental conditions over which a $T^2$ dependence of the carrier mobility has been reported~\cite{Mikheev15APL,Klimin12,Lin15}.  STO has a low critical density for degenerate doping~\cite{Spinelli10}, so we assume that all the electron donors remain ionized as a function of temperature, leading to a constant carrier density.  We also assume the cubic structure and neglect spin-orbit coupling.  Away from the conduction-band minimum at the $\Gamma$ point, the bands split into two ``light'' bands (mass $m_\pi$) and one ``heavy'' band (mass $m_\delta$) with lobes along the Cartesian directions.  This allows momentum relaxation through scattering between bands.  We model the dispersion relation as an anisotropic parabola:
\begin{equation}
\eps_i =\frac{\hbar^2k_x^2}{2m_{ix}}+\frac{\hbar^2k_y^2}{2m_{iy}}+\frac{\hbar^2k_z^2}{2m_{iz}} \, ,
\label{STO_dispersion}
\end{equation}
where $i$ indexes the three conduction bands, $m_{1x}=m_{2y}=m_{3z}=m_\delta$, and the other masses are $m_\pi$.
The effective electron-electron interaction as derived in Ref.~\onlinecite{Klimin12} is
\begin{equation}
U_{\vq}^\text{(eff)} = \frac{4\pi e^2}{\eps_\infty (q^2+\kappa^2)} - \sum_{\lambda}\frac{2\abs{V_{\vq,\lambda}}^2}{\hbar\omega_\lambda(q^2+\kappa^2)^2}- \frac{2\abs{V_\vq^{(ac)}}^2}{\hbar \omega_\vq^{(ac)}} \, .
\end{equation}
The three terms represent screened Coulomb interaction, optical-phonon-mediated scattering, and acoustic-phonon-mediated scattering.  $\eps_\infty$ is the high-frequency dielectric constant, $\kappa^2$ is the Lindhard screening length defined above, and $V_{\vq,\lambda}$ is the Fr{\"o}hlich interaction with optical phonons~\cite{Devreese10}.  The acoustic potential is given by
\begin{equation}
V_\vq^{(ac)} = \sqrt{4\pi \alpha_{ac}} \frac{\hbar^2}{m_D}q^{1/2} \quad  \text{with} \quad \alpha_{ac} = \frac{E_d^2 m_D^2}{8\pi n \hbar^3 v} \, ,
\end{equation}
where $\omega_\vq = v q$ is the acoustic phonon frequency, $n=5.11\, \text{g}\, \text{cm}^{-3}$ is the density of STO, $E_d$ is the deformation potential, $m_D=(m_\pi^2m_\delta)^{1/3}$ is the density-of-states mass, and $v=8.1\times10^3\, \text{m/s}$ is the speed of sound in STO.

Starting from a first-principles band structure (using {\sc VASP} with the Heyd, Scuseria, and Ernzerhof~\cite{HSE03,HSE06} functional), we fit $m_\pi = 1.16 m_e$ and $m_\delta = 15.31 m_e$, with an accuracy better than 8 meV in the region of interest.  Optical phonon frequencies were taken from the calculations in Ref.~\onlinecite{Himmetoglu14}, and the deformation potential for the conduction band was taken to be $-4.0$ eV as calculated in Ref.~\onlinecite{Janotti12}.

\begin{figure}
\includegraphics[width=8.6cm]{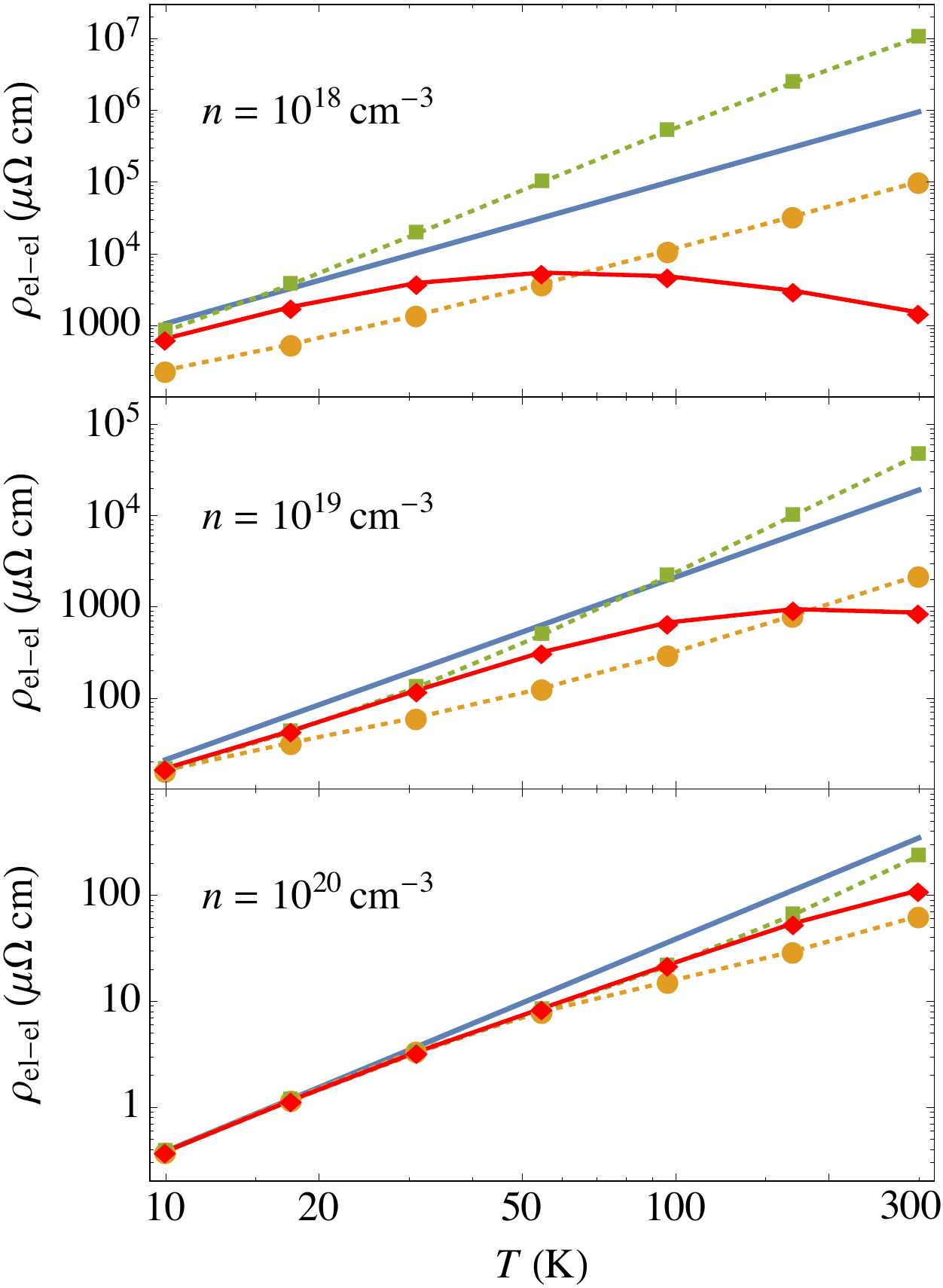}
\caption{Log-log plots of $\rho_\text{el-el}$ versus $T$ in STO, with zero-temperature carrier densities as indicated in the panels.  Solid blue lines are the pure $T^2$ law, obtained using all three Assumptions.  Orange circles are calculated by computing $I$ [Eq.~(\ref{I})] numerically (i.e., by removing Assumption 3).  Green squares show data for which the radial variation of the non-Fermi-function terms is also included (removing Assumptions 2 and 3).  Red diamonds show data in which the chemical potential is also allowed to move with temperature to keep the carrier density fixed; all the Assumptions have been removed, so this represents our final results for the electron-electron resistivity.}
\label{STO}
\end{figure}

With all Assumptions in place, $\rho_\text{el-el}\propto T^2$, as shown in Fig.~\ref{STO}.  However, none of these Assumptions actually hold in STO, due to the significant change of the chemical potential with temperature and its position close to the band edge, as illustrated in Fig.~\ref{muvsT}.  The full result, obtained without any assumptions, does not follow a $T^2$ power law (Fig.~\ref{STO}).  The deviation is particularly pronounced for lower doping ($10^{18}$ cm$^{-3}$).  The temperature dependence is closer to $T^2$ for higher doping ($10^{20}$ cm$^{-3}$) because the chemical potential is higher and changes less with temperature, so the Assumptions are closer to being satisfied.

It is worthwhile to discuss our results in the context of earlier work by Klimin \emph{et al.} \cite{Klimin12}.  While we use the same expression for $\rho_\text{el-el}$, our values for the parameters are different and thus we obtain a different scattering rate.  Our masses give a better fit to the first-principles band structure of STO, and our deformation potential is calculated instead of being used as a fitting parameter.
This affects the relative contributions of the various scattering mechanisms:
while we find that Coulomb scattering is dominant, Klimin \emph{et al.} found a competition between Coulomb and acoustic-phonon-mediated scattering due to their much larger deformation potential (23.3 eV).  However, their $\rho_\text{el-el}\propto T^2$ dependence is a result of employing the Assumptions, so these quantitative differences do not impact our main conclusion.  Tests employing the assumptions and parameters used by Klimin \emph{et al.} \cite{Klimin12} reproduce their results, and if we use their parameters but do not make the Assumptions, no $T^2$ dependence is found.


In summary, numerical calculations of electron-electron scattering in SrTiO$_3$ do not show a $T^2$ power law.  This deviates from the typical Baber scattering result because several key assumptions used to derive the $T^2$ exponent are not satisfied in the SrTiO$_3$ system.  Our case study illustrates that electron-electron scattering does not always lead to a $T^2$ power law, particularly in systems in which the chemical potential of the electrons may be close to a band edge.  This result shows that caution must be used when attempting to identify the physical mechanism behind an observed $T^2$ power law.

\begin{acknowledgments}
We acknowledge B. Himmetoglu, K. Krishnaswamy, J.-X. Shen, D.\ Wickramaratne, E. Mikheev, S. Stemmer, L. Balents, and S. J. Allen for fruitful discussions, and especially J.\ T.\ Devreese for invaluable correspondence.
This work was supported by the MURI program of the Office of Naval Research, Grant No. N00014-12-1-0976.
Computational resources were provided by the Extreme Science and Engineering Discovery Environment (XSEDE), which is supported by National Science Foundation (NSF) grant number ACI-1053575, and by the Center for Scientific Computing from the CNSI, MRL: an NSF MRSEC (DMR-1121053) and NSF CNS-0960316.
\end{acknowledgments}

\bibliography{el-el}

\end{document}